\documentclass{article}
\usepackage[utf8]{inputenc}
\usepackage{hyperref}
\usepackage{xcolor}
\usepackage{graphicx}
\usepackage{textcomp}

\begin{document}

\title{Over 20-year global magnetohydrodynamic simulation of Earth's magnetosphere}
\author{Ilja Honkonen, Max van de Kamp, Theresa Hoppe, Kirsti Kauristie\\Finnish Meteorological Institute, POB 503, 00101 HELSINKI, Finland\\firstname.lastname@fmi.fi}
\date{\today}
\maketitle

\abstract
We present our approach to modeling over 20 years of the solar wind-magnetosphere-ionosphere system using version 5 of the Grand Unified Magnetosphere-Ionosphere Coupling Simulation (GUMICS-5).
As input we use 16-s resolution magnetic field and 1-min plasma measurements by the Advanced Composition Explorer (ACE) satellite from 1998 to 2020.
The modeled interval is divided into 28 h simulations, which include 4 h overlap.
We use a maximum magnetospheric resolution of 0.5 Earth radii (R$_E$) up to about 15 R$_E$ from Earth and decreasing resolution further away.
In the ionosphere we use a maximum resolution of approximately 100 km poleward of $\pm58^o$ magnetic latitude and decreasing resolution towards the equator.
With respect to the previous version GUMICS-4, we have parallelized the magnetosphere of GUMICS-5 using the Message Passing Interface and have made several improvements which have e.g. decreased its numerical diffusion.

In total we have performed over 8000 simulations which have produced over 10 000 000 ionospheric files and 2 000 000 magnetospheric files requiring over 100 TB ($10^{14}$ bytes) of disk space.
We describe the challenges encountered when working with such a large number of simulations and large data set in a shared supercomputer environment.
Automating most tasks of preparing, running and archiving simulations as well as post-processing the results is essential.
At this scale the available disk space becomes a larger bottleneck than the available processing capacity, since ideally the simulations are stored permanently but only have to be executed once.

We compare the simulation results to several empirical models and geomagnetic indices derived from ground magnetic field measurements.
GUMICS-5 reproduces observed solar cycle trends in magnetopause stand-off distance and magnetospheric lobe field strength but consistency in plasma sheet pressure and ionospheric cross-polar cap potential is lower.
Comparisons with geomagnetic indices show better results for $Kp$ index than for $A_E$ index.

The simulation results are available at \href{https://doi.org/10.23729/ca1da110-2d4e-45c4-8876-57210fbb0b0d}{doi.org/10.23729/ca1da110-2d4e-45c4-8876-57210fbb0b0d}, consisting of full ionospheric files and size-optimized magnetospheric files.
The data used for Figures is available at \href{https://doi.org/10.5281/zenodo.6641258}{doi.org/10.5281/zenodo.6641258}.

Our extensive results can serve e.g.~as a foundation for a combined physics-based and black-box approach to real-time prediction of near-Earth space, or as input to other physics-based models of the inner magnetosphere, upper and middle atmosphere, etc.

\section{Introduction}

Global magnetohydrodynamic (GMHD) simulations have been used for studying the interaction between solar wind, magnetosphere and ionosphere since the late 70s and early 80s (e.g.~\cite{leboeuf81, brecht81, wu81} and references therein) and have become an especially important tool for understanding strong and/or dynamic events for which e.g.~empirical models are either not suitable or lack observations.
Models based on first principles, such as GMHD, might also help understand space climate - the long-term effects of the Sun on the Earth's magnetosphere, ionosphere and thermosphere - but currently the ability and accuracy of GMHD models in reproducing space weather over time scales of one or more solar cycles is unknown.

So far GMHD models have been mostly used and validated for individual events and output parameters or synthetic solar wind inputs.
Continuous simulations of more than one week are rare and in many cases the simulation outputs are not available publicly.
Table \ref{tab:sims} lists the longest published GMHD simulations to date which use real solar wind as input and the domain(s) or parameter(s) that were studied.
In most cases long runs have been used in statistical validation of GMHD performance.
Studies listed in Table \ref{tab:sims} report on GMHD models' ability to forecast geomagnetic activity \cite{haiducek17, liemohn18,camporeale20}, to map between magnetosphere and ionosphere along magnetic field lines \cite{facsko16}, and to reproduce observed magnetospheric \cite{guild08, guild08b} or ionospheric properties \cite{zhang11, wiltberger17}.
\cite{juusola14} compares seasonal variability in the observed modified $A_E$ index with that of a one-year run produced by GUMICS-4.

\begin{table}
\label{tab:sims}
\caption{Longest published global magnetohydrodynamic simulations performed to date.}
\begin{tabular}[c]{l|c|p{4cm}|p{2cm}}
\hline
Interval (days) & Model & Subjects & Reference(s) \\
\hline
\hline
2015-04-19/2017-07-17 (821) & SWMF & Dst, ground $dB/dt$ & \cite{liemohn18} \cite{camporeale20} \\
\hline
2002-01-30/2003-02-02 (369) & GUMICS-4 & ionosphere, $\vec{B}$ mapping, plasma sheet & \cite{juusola14} \cite{kallio15} \cite{facsko16} \\
\hline
1996-02-23/1996-04-26 (64) & LFM  & ionosphere, plasma sheet & \cite{zhang11} \cite{guild08} \cite{guild08b} \\
\hline
2005-01-01/2005-01-31 (31) & SWMF & $Kp$, {\it SYM-H}, $A_L$, {\it CPCP} & \cite{haiducek17} \\
\hline
2008-03-20/2008-04-16 (28) & LFM & {\it FAC}, {\it CPCP} & \cite{wiltberger17} \\
\hline
1998-02-06/2020-12-31 ($\approx$8300) & GUMICS-5 & magnetopause, magnetosphere, plasma sheet, ionosphere, geomagnetic indices & this work \\
\hline
\end{tabular}
\end{table}

We present the first results from GMHD simulations that span several solar cycles - from 1998 to 2020.
Our main objective is to determine whether GUMICS-5 can reproduce the time evolution of magnetosphere and ionosphere over such time scales.
Section \ref{sec:setup} describes our model, input parameters and the challenges associated with performing over 8000 1-day simulations and post-processing the results.
In section \ref{sec:results} we compare the results to empirical models of the ionosphere and magnetosphere, and to measured indices of the Earth's geomagnetic field.
We discuss the results in section \ref{sec:discussion} and draw our conclusions in section \ref{sec:conclusions}.

\section{Modeling and post-processing setup}
\label{sec:setup}

\subsection{GUMICS-5}

We use the fifth major version of Grand Unified Magnetosphere-Ionosphere Coupling Simulation (GUMICS-5), which is a parallelized version of GUMICS-4 described extensively in \cite{janhunen12}.

In the magnetosphere GUMICS-4 and 5 solve the ideal MHD equations using a first-order conservative finite volume method.
The MHD equations are solved primarily using Roe's approximate Riemann solver \cite{roe81} and in rare cases when that solution fails a more robust and numerically diffusive solver is used for that particular calculation.
Similarly to GUMICS-4 and \cite{tanaka94}, the magnetic field is split into a static background dipole ($B_0$) and a perturbed part ($B_1$) which is modified by the MHD solver(s).
The ionosphere is modeled as a spherical surface with radially integrated current continuity, from which the electric potential is solved using a dipole-field-aligned electric current and a radially integrated conductivity tensor. Ionospheric conductivity is controlled by the magnetospheric temperature and mass density and by a parameter characterizing the loss cone filling rate. The ionosphere - thermosphere interaction processes impacting conductivities are based on the MSIS model \cite{hedin91}.

On a high level the ionospheric and magnetospheric grids of GUMICS-5 are identical to GUMICS-4:
The magnetosphere is discretized as a cell-by-cell run-time adaptive octogrid and in GUMICS-5 it is built on a custom version of DCCRG \cite{honkonen13} with the solution calculated in parallel using the Message Passing Interface.
The ionosphere is modeled as a spherical surface and is discretized as a static cell-by-cell adaptive triangular grid.
In GUMICS-4 the ionospheric grid had a maximum resolution of approximately 100 km only in the auroral oval region, while in GUMICS-5 the maximum resolution has been extended to everywhere poleward of $\pm58^o$ magnetic latitude.

The initial objective for developing GUMICS-5 was to obtain identical results compared to GUMICS-4 with significantly faster time to solution.
During subsequent development several features of the model were improved, e.g.~removal of divergence of perturbed magnetic field and mapping between ionosphere and magnetosphere along dipole field lines, and the code was updated to a more recent C++ standard used by current compilers.
Perhaps most importantly, a rotating background dipole field was added that was not available in GUMICS-4.
With these modifications, results between GUMICS-4 and -5 are very similar overall but not identical.

\subsection{Simulation parameters}

The interval 1998-2020 is divided into simulations of 1 UTC day with an additional 4 h overlap with the previous day's simulation. A 4 h overlap was considered as a safe value, since \cite{facsko16} showed that, along the orbit(s) of Cluster spacecraft, switching from one simulation to another after 1 h of overlap produced on average a discontinuity smaller than the natural variation of the results.
As shown in Section \ref{sec:convergence} a 4 h overlap is indeed sufficient for emulating one continuous simulation using several independent shorter simulations.

\begin{figure}
\centerline{\includegraphics[width=0.5\paperwidth]{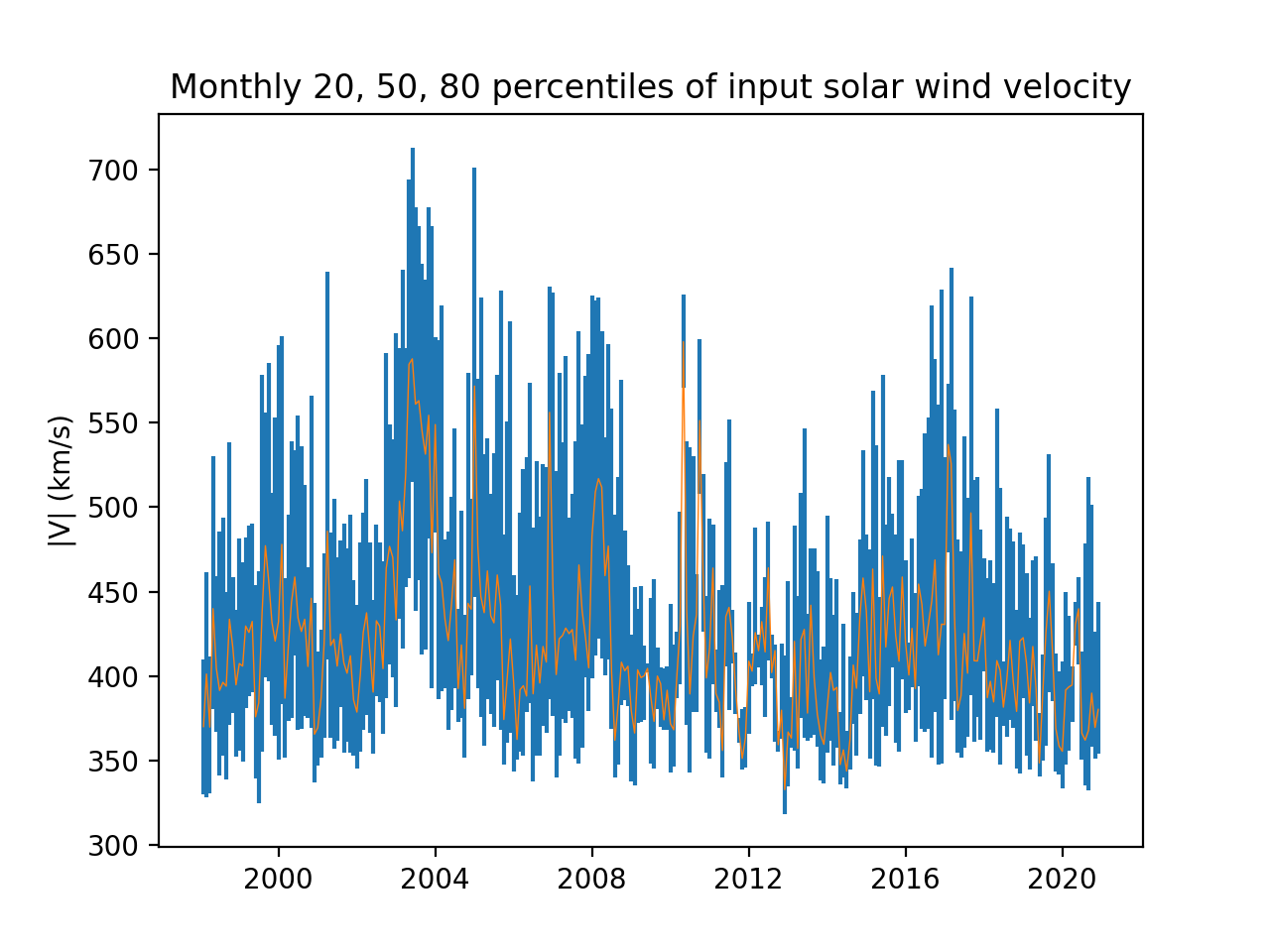}}
\caption{
Monthly 20, 50 and 80 percentiles of solar wind velocity magnitude calculated from input given to GUMICS-5.
Median is marked by solid line and vertical bars show the range of 20 and 80 percentiles.
}
\label{fig:ace}
\end{figure}

As input we use solar wind data measured by Advanced Composition Explorer (ACE) consisting of 1-min resolution plasma data and 16-s magnetic field data, available under 
\href{https://cdaweb.gsfc.nasa.gov/pub/data/ace}{cdaweb.gsfc.nasa.gov/pub/data/ace}.
Missing and invalid data are interpolated linearly from the nearest existing data regardless of gap length, except for gaps that spanned a change in calendar year.
Because of this, approximately 20 days of simulations were not performed due to missing solar wind data, often during the last few days of December on several years.
Figure \ref{fig:ace} shows monthly 20, 50 and 80 percentiles of the solar wind velocity given to GUMICS-5 as input, with the median marked by a solid line and vertical bars showing the range of 20 and 80 percentiles.

The simulation extends to $\pm 248 R_E$ in X dimension in geocentric solar ecliptic (GSE) coordinates, where $R_E = 6371$ km is the radius of Earth.
In GSE Y and Z dimensions the simulation extends to $\pm 64 R_E$.
We use a maximum magnetospheric resolution of 0.5 Earth radii ($R_E$) up to about 15 $R_E$ from Earth and a decreasing resolution further away down to a minimum resolution of 8 $R_E$.
The boundary of the simulation on the Sun-ward side is set every minute to the position of ACE although it rarely changes from one layer of grid cells to another during one simulation.
With the correct location for the source of solar wind data to within approximately $4 R_E$, the simulation should give realistic timings for modeled phenomena within limitations of the solved MHD and electrostatic equations.

The background dipole field direction is updated every minute, the divergence of the perturbed magnetic field is removed every 20 s and the ionospheric potential solution is updated every 4 s.
We save snapshots of the ionospheric state every simulated minute and of the magnetospheric state every 5 minutes, resulting in approximately 500\ 000 and 100\ 000 files per year respectively.

\subsection{Challenges of scale}

Most of the previously manual steps of using a GMHD model had to be automated because of the scale of the endeavor, as nearly 8000 simulations were prepared, executed, post-processed and transferred to long-term storage.
In total over 2\ 000\ 000 magnetospheric output files and 10\ 000\ 000 ionospheric output files were produced with a total size of over 100 TB ($10^{14}$ bytes).

We use Finnish Meteorological Institute's (FMI) partition on the Puhti supercomputer operated by the Finnish Center for Scientific Computing.
The theoretical peak performance of Puhti is 1.8 petaflops using 682 nodes of which 238 belong to the partition dedicated for FMI.
Puhti's Lustre filesystem has 4.8 petabytes of space of which 0.6 PB is dedicated for FMI.
The resources of FMI partition are shared between all users at FMI which resulted in some complications.

In practice disk space was always limited, despite a total size of almost 600 TB, and allowed the processing of only 2-4 years of simulation results simultaneously, with the rest having to be kept on a separate system designed for long-term storage.
Also the computational load of the system varied widely, usually from almost full during business hours to almost empty during weekends and holidays.

In order to take full advantage of available computational resources without interfering with other users, a simulation scheduling program was developed.
The scheduler was executed every hour to keep 20 simulations constantly running or queuing, using almost 10 \% of total computational capacity, and to additionally start as many simulations as would fit into the FMI partition during weekends.
Often the scheduling program was also run manually to start additional simulations in case significant computational capacity was available.
We estimate that taking advantage of additional capacity available during weekends corresponded to constantly running 40-60 simulations representing 15-25 \% of total capacity of FMI's partition.

We solved limitations of disk space during post-processing by implementing a parallel wrapper program which downloaded results from long-term storage transparently and on-demand, and subsequently executed multiple copies of the specified serial program for post-processing results in parallel.
By default this also included verifying checksums of downloaded files.
A separate program was developed for initially calculating checksums, uploading the results and verifying that the upload succeeded by downloading the results into a temporary directory for verification.
Deletion of results from local disk after sufficient post-processing was left for the user, as the manual steps are trivial.

While most tasks related to GUMICS-5 were automated, some were left to be done manually as they were either infrequent enough or too laborious to automate at the time.
For example several times the local disk became full due to other users, meaning that most simulations and post-processing failed and had to be cleaned up and restarted.
Latest version of the scheduling program is also able to restart such failed runs.
On many occasions data transfers from permanent storage to local supercomputer disk failed silently or slowed down enough to require manual intervention.
These cases might also be handled automatically in the future versions of data transfer and post-processing programs.
Simulations also "failed" on approximately 20 days when solar wind data was not available for the entire 28 h interval.

\section{Results}
\label{sec:results}

We compare the magnetospheric and ionospheric results to empirical models of the magnetosphere and ionosphere, similarly to \cite{gordeev13, gordeev15}.
This avoids many of the challenges related to interpreting data from a diverse collection of space instruments located in different parts of the magnetosphere at different stages of the solar cycle or even different solar cycles.
These challenges have been addressed adequately by the developers of said empirical models, making them suitable for comparison with the results of GUMICS-5 presented here.

As solar wind input for empirical models, we use the output of GUMICS-5 taken at $X=12R_E, Y=44R_E, Z=44R_E$ in GSE coordinates.
This is well outside of the modeled magnetosphere regardless of the location of the bowshock and represents the undisturbed solar wind impacting the magnetosphere.
As demonstrated in \cite{claudepierre10}, GMHD models act as a low-pass filter for solar wind, with the cutoff frequency determined mainly by magnetospheric resolution and the solar wind speed.
Using the solar wind modeled by GUMICS thus avoids the highest-frequency variations from appearing in empirical results which would consist e.g.~of unrealistically fast signal propagation through the magnetosphere and ionosphere.
On a monthly scale the solar wind as modeled by GUMICS-5 (not shown) is nearly identical to the observations in Figure \ref{fig:ace}.

The use of over 20 years of real, albeit low-pass filtered, solar wind data comprises a significant difference between our comparison and that of e.g.~\cite{gordeev15}, where approximately 200 hours of simulations were run with quasi-stationary solar wind data.

In the heatmaps below, color coding indicates the number of data points in each bin. The total number of data points is approximately 100\ 000 for magnetospheric results and 500\ 000 for ionospheric results, per year.
We present an overview of our results for the years 1998-2020 and show details for the maximum of solar cycle 23 in 2001 and for the subsequent minimum in 2009.

\subsection{Convergence of overlapping simulations}
\label{sec:convergence}

\begin{figure}
\centerline{\includegraphics[width=0.7\paperwidth]{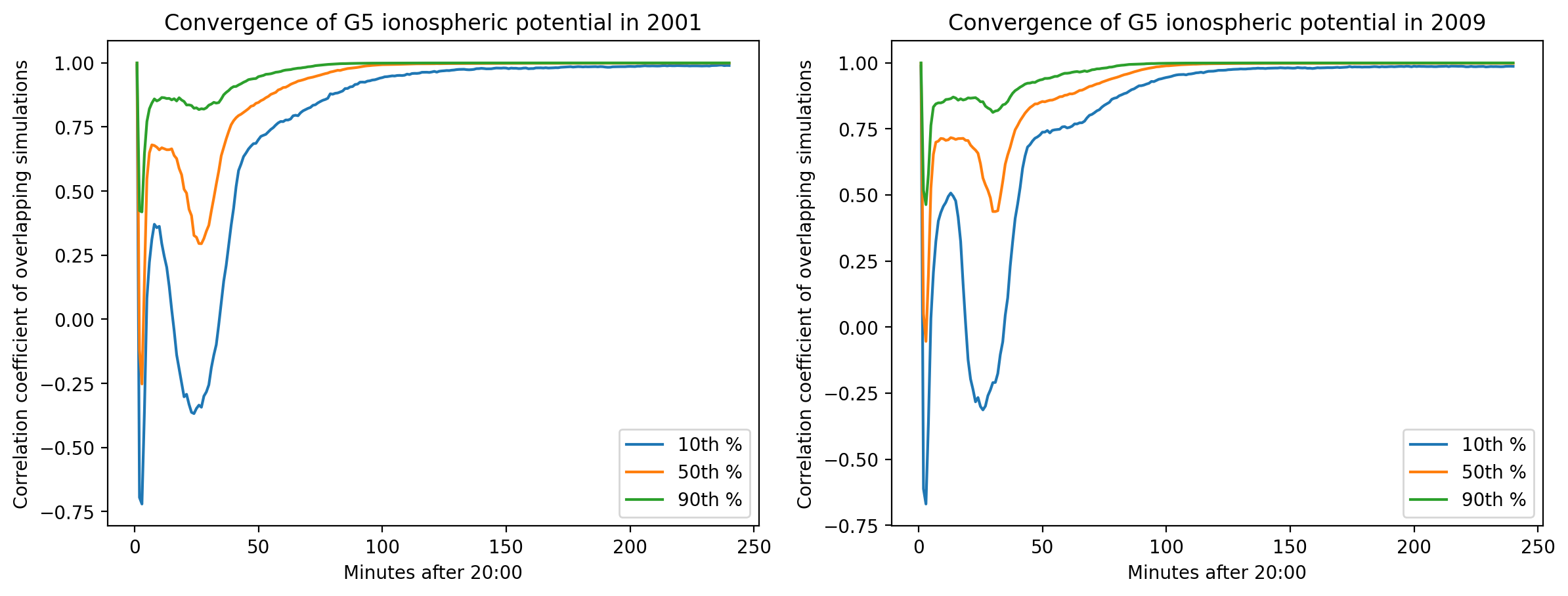}}
\caption{
Convergence of GUMICS-5 ionospheric potential during maximum of solar cycle 23 (2001) and the following minimum (2009).
The total number of data points at each minute after 20:00 UTC in each plot is approximately 364.
}
\label{fig:iconv}
\end{figure}

In order to test the convergence of GUMICS-5 results, Figure \ref{fig:iconv} shows the correlation coefficients ($CC$) between ionospheric potentials of overlapping simulations for years 2001 and 2009.
For each point in time that was modeled by two simulations, the $CC$ is calculated between electric potentials of all ionospheric cells of both simulations.
Thus for each minute between 20:00 and 23:59 we obtain up to 364 $CC$s each calendar year.

The 10th, 50th, and 90th percentiles of $CC$s are shown as functions of the number of minutes since beginning of the overlaps.
The convergence results for other years are very similar to those shown in Figure \ref{fig:iconv}.

The ionospheric results converge after about 3 h of overlap.
Even after 2 h of overlap, using separate simulations instead of one continuous simulation most likely results in smaller errors than from other sources discussed in Section \ref{sec:errors}.
As shown in \cite{facsko16}, the dayside magnetosphere converges sufficiently even in 1 h and we assume that the magnetosphere converges faster than the ionosphere, hence the 4 h overlap used here should guarantee that the error from switching from one simulation to the next is insignificant compared to other sources.
After about 1 h of overlap, convergence in the ionosphere appears to be very similar regardless of the phase of solar cycle.

\subsection{Comparison to higher resolution GUMICS-4 one year run}

In order to assess the differences to previous version of GUMICS, we compare our results to those presented in \cite{juusola14} which were obtained with the previous single-core version GUMICS-4, using a twice higher maximum spatial resolution of 0.25 R$_E$ in the magnetosphere (\cite{facsko16}).

\subsubsection{Modified $A_L$ and $A_U$ indices}

\begin{figure}
\centerline{\includegraphics[width=0.5\paperwidth]{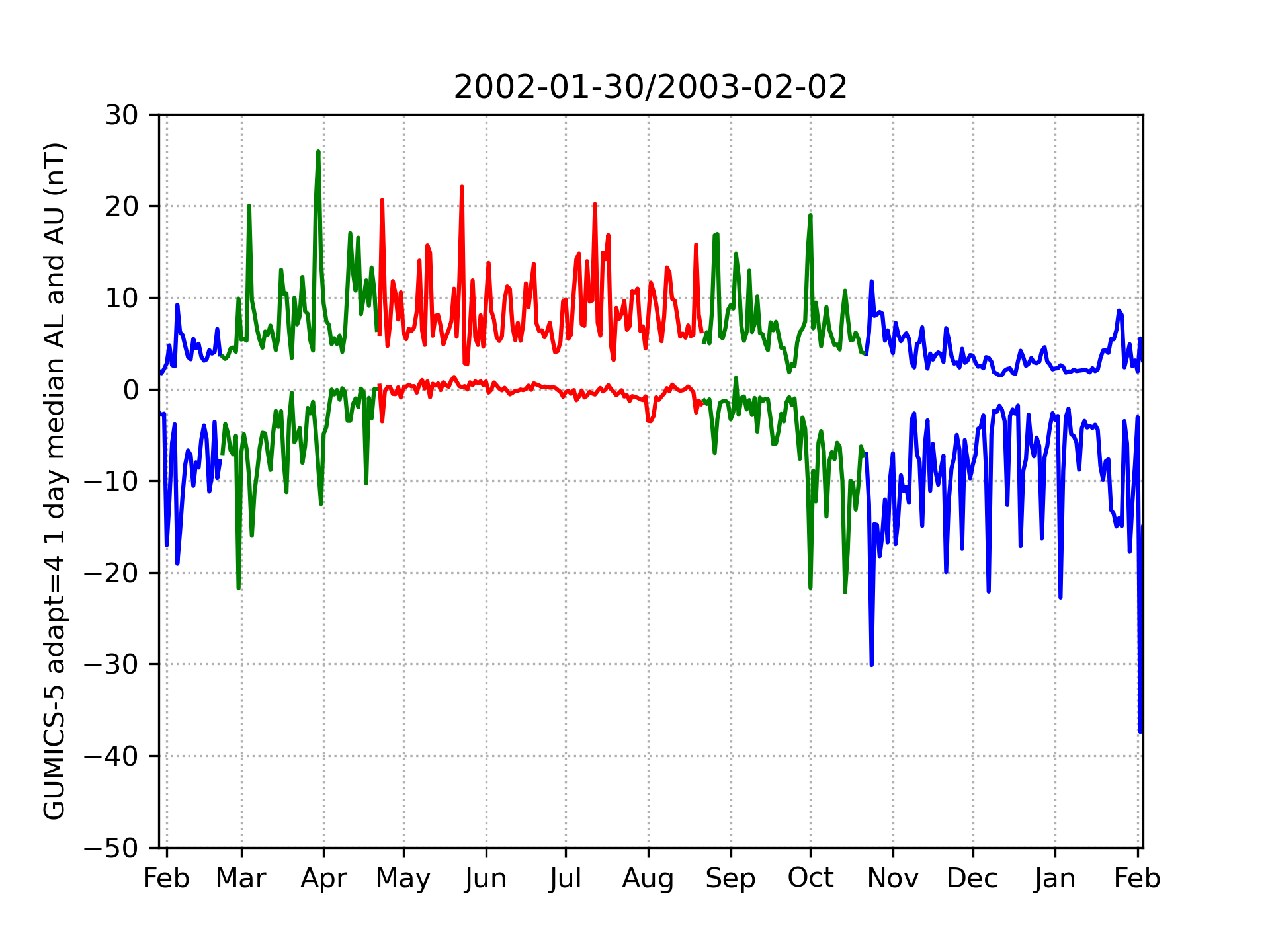}}
\caption{
Daily median modified $A_L$ and $A_U$ indices in same format as Figure 4a of \cite{juusola14} but produced using GUMICS-5 and highest magnetospheric resolution of 0.5 R$_E$ instead of GUMICS-4 and highest resolution of 0.25 R$_E$.
}
\label{fig:yrrunae}
\end{figure}

Figure \ref{fig:yrrunae} shows the daily median modified $A_L$ and $A_U$ indices produced with GUMICS-5, in the same format as Figure 4a of \cite{juusola14}.
The $A_L$ and $A_U$ indices are described in \cite{davis66} and characterize time variations in the intensity of westward and eastward auroral electrojet currents.
Instead of the standard derivation of $A_U$ and $A_L$ indices, we use the procedure from \cite{juusola14}:
The modified indices are computed as minimum and maximum of the northward components of external ground magnetic field at all $A_E$ stations.

An exact one-to-one match between GUMICS-4 and 5 cannot be expected due to additional features and improvements of GUMICS-5 but the overall behavior of daily median modified $A_L$ and $A_U$ indices are very similar.
The GUMICS-based indices have a seasonal dependence that does not exist in observations.
Both versions of the code underestimate the absolute values of modified $A_L$ and $A_U$ compared to observations.
Underestimation of $A_L$ is particularly strong in the summer and of $A_U$ in winter.

\begin{figure}
\centerline{\includegraphics[width=0.6\paperwidth]{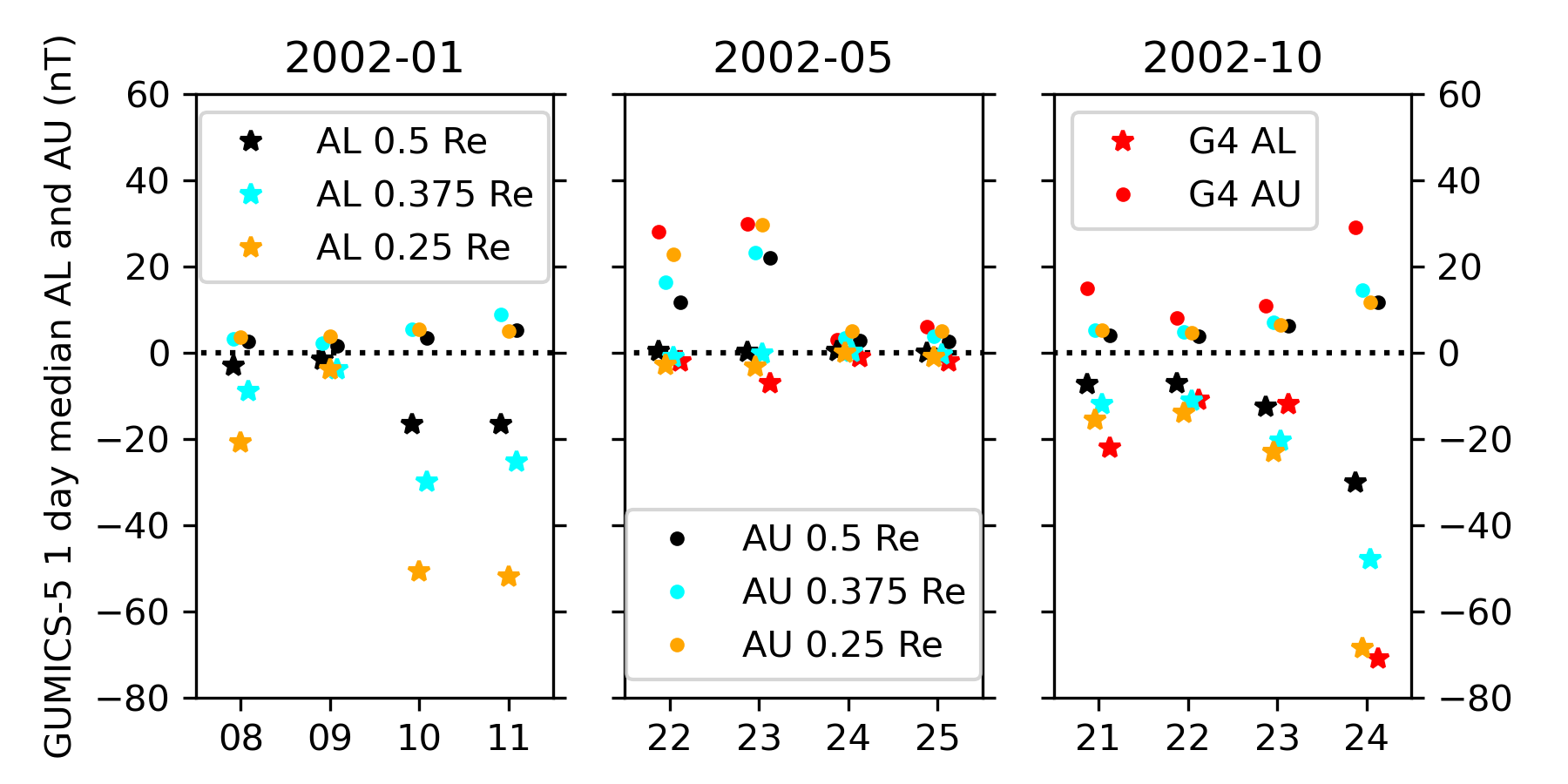}}
\caption{
Comparison of modified daily median $A_L$ and $A_U$ indices from GUMICS-5 using highest magnetospheric resolution of 0.5, 0.375 and 0.25 R$_E$ and from GUMICS-4 using highest magnetospheric resolution of 0.25 R$_E$ (\cite{juusola14}).
Note that a small horizontal offsets has been applied for clarity and that GUMICS-4 values are approximate.
}
\label{fig:adaptcomp}
\end{figure}

Comparison of Figure \ref{fig:yrrunae} with Figure 4a in \cite{juusola14} reveals that the extreme values of $A_U$ and $A_L$ of GUMICS-5 are systematically smaller than those of GUMICS-4.
This discrepancy is mainly the result of different magnetospheric resolutions used in the two simulations:
Figure \ref{fig:adaptcomp} shows daily median modified $A_L$ and $A_U$ indices produced with GUMICS-5 on 12 selected days using highest magnetospheric resolutions of 0.5, 0.375 and 0.25 $R_E$, and GUMICS-4 using highest magnetospheric resolution of 0.25 R$_E$.
The absolute values of daily median $A_U$ and $A_L$ increase systematically with magnetospheric resolution used in the simulation.
When using a resolution of 0.25 $R_E$, the magnitudes of $A_U$ and $A_L$ are consistent between both GUMICS-5 and 4.
For example, both versions yield a daily median modified $A_L$ of approximately -70 nT on 2002-10-24 and $A_U$ of approximately 30 nT on 2002-05-23.

Overall the daily median modified $A_U$ index produced by GUMICS-5 is similar to that of GUMICS-4.
For example, the peaks in the beginning and end of 2002-03, as well as the peaks in 2002-10, are local maximums in both versions with similar relative amplitudes.

The $A_L$ index produced by GUMICS-5 is very similar to that of GUMICS-4, for example, from 2002-10 until the end of the modeled interval.
A few notable differences are e.g.~a missing negative peak in 2002-11-01 in GUMICS-5, and that the peak of 2002-11-21 is smaller than the peak of 2002-12-07 in GUMICS-5 (-20 vs -22 nT), while being larger in GUMICS-4 (-70 vs -50 nT).

Together Figures \ref{fig:yrrunae} and \ref{fig:adaptcomp} strongly suggest that in this work GUMICS-5 can be safely used in place of GUMICS-4.

\subsubsection{Cross-polar cap potential in northern hemisphere}

\begin{figure}
\centerline{\includegraphics[width=0.5\paperwidth]{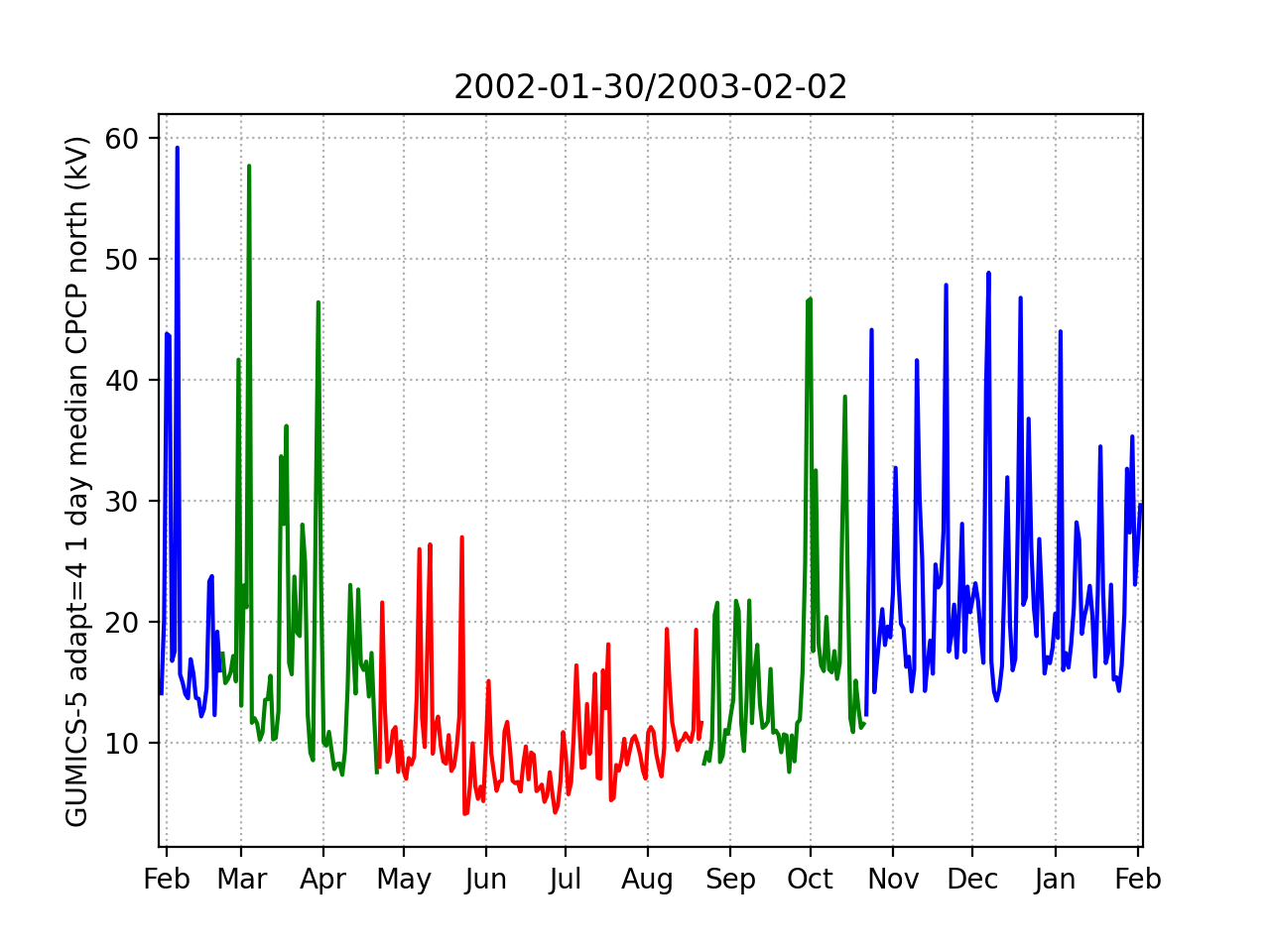}}
\caption{
Daily median cross-polar cap potential in northern hemisphere in same format as Figure 1a of \cite{juusola14} but produced using GUMICS-5 and highest magnetospheric resolution of 0.5 R$_E$ instead of GUMICS-4 and highest resolution of 0.25 R$_E$.
}
\label{fig:cpcpndaily}
\end{figure}

Figure \ref{fig:cpcpndaily} shows the daily median cross-polar cap potential (CPCP) in the northern hemisphere from GUMICS-5 in the same format as Figure 1a of \cite{juusola14} from GUMICS-4.
The overall behavior and many details of GUMICS-5 solution are same as in GUMICS-4 but there are also more differences between the models than with modified $A_L$ and $A_U$ indices.
During the northern summer months, the GUMICS-5 daily median CPCP is somewhat smaller than in GUMICS-4, but during winter the difference is approximately a factor of two, similarly to $A_L$ and $A_U$ indices. 

\begin{figure}
\centerline{\includegraphics[width=0.5\paperwidth]{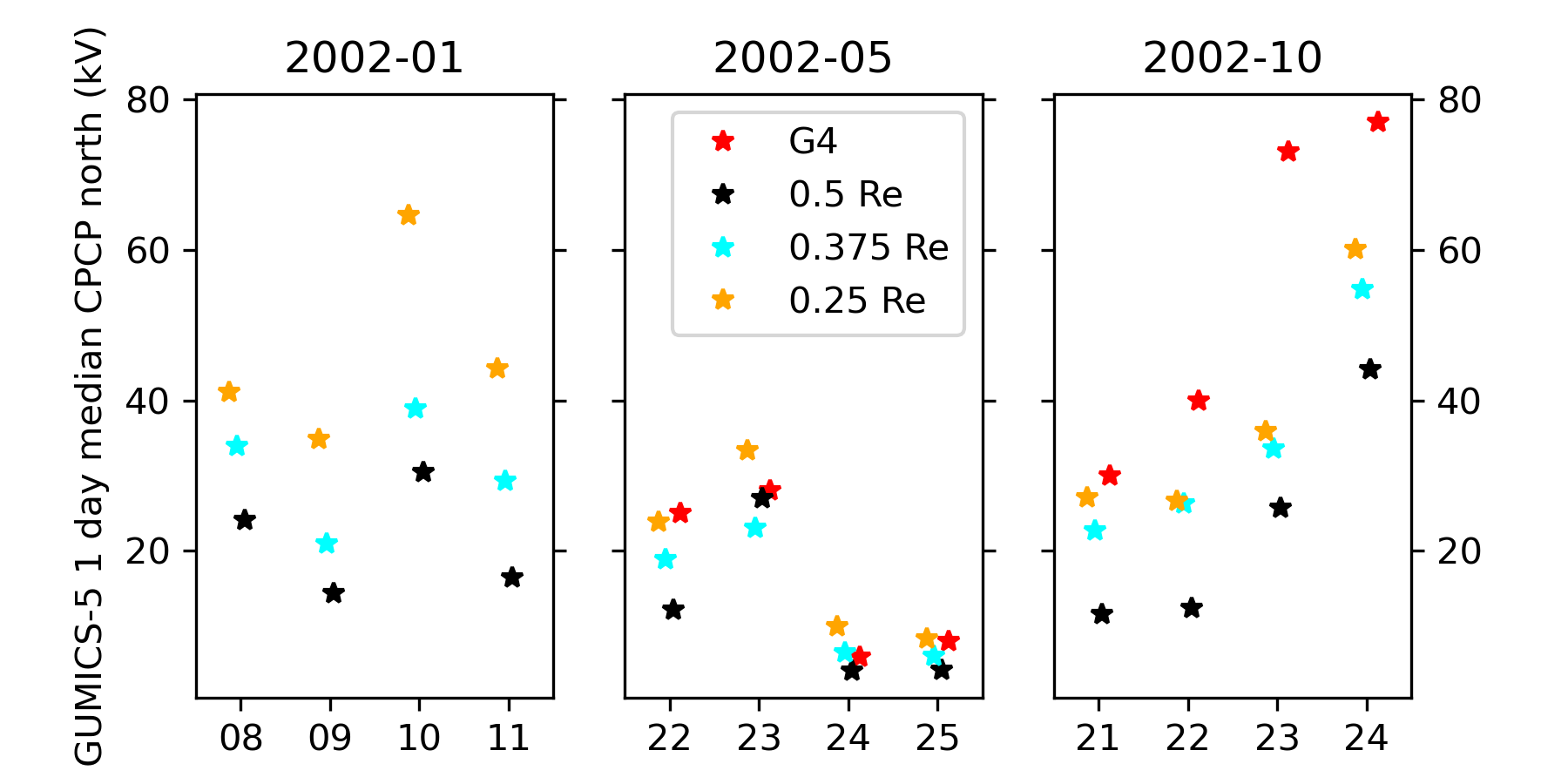}}
\caption{
Comparison of daily median cross-polar cap potential in northern hemisphere from GUMICS-5 using highest magnetospheric resolution of 0.5, 0.375 and 0.25 R$_E$, and from GUMICS-4 (\cite{juusola14}) using highest magnetospheric resolution of 0.25 R$_E$.
Note that a small horizontal offsets has been applied for clarity and that GUMICS-4 values are approximate.
}
\label{fig:cpcpzoom}
\end{figure}

Figure \ref{fig:cpcpzoom} shows north CPCP produced with GUMICS-5 on select days using highest magnetospheric resolutions of 0.5, 0.375 and 0.25 $R_E$.
The effect of increasing magnetospheric resolution is not as straightforward in CPCP as in $A_L$ and $A_U$.
For example on 2002-01-10, CPCP increases from 30 kV at 0.5 $R_E$ to over 60 kV at 0.25 $R_E$ while on 2002-10-23 CPCP increases only from 25 to 35 kV.
The artificial seasonal variation of CPCP seems weaker in GUMICS-5 than in GUMICS-4 but this might be due to lower magnetospheric resolution.

Based on comparisons of modified $A_E$ and CPCP we conclude that GUMICS-5 can be used in place of GUMICS-4 at least on solar cycle time scales.
It should be noted that, although in general using higher magnetospheric resolution should improve statistical consistency between simulations and observations, this does not necessarily hold in individual cases.
Also as \cite{ridley10} showed, even using the highest feasible resolution in one short simulation may not guarantee that the results converge to any particular value.

\subsection{Magnetospheric lobe field strength}

\begin{figure}
\centerline{\includegraphics[width=0.9\paperwidth]{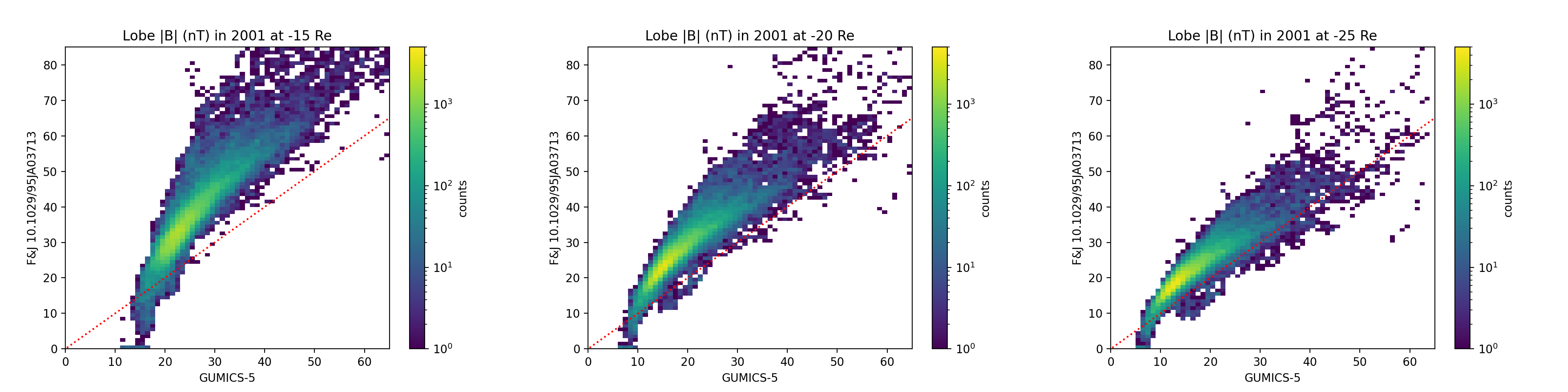}}
\caption{
Magnetospheric lobe field strength during the maximum of solar cycle 23 at -15, -20 and -25 R$_E$ GSE X in Earth's magnetotail.
The red line indicates 1:1 correspondence.
The total number of data points in each plot is approximately $10^5$.
}
\label{fig:lobe2001}
\end{figure}

\begin{figure}
\centerline{\includegraphics[width=0.9\paperwidth]{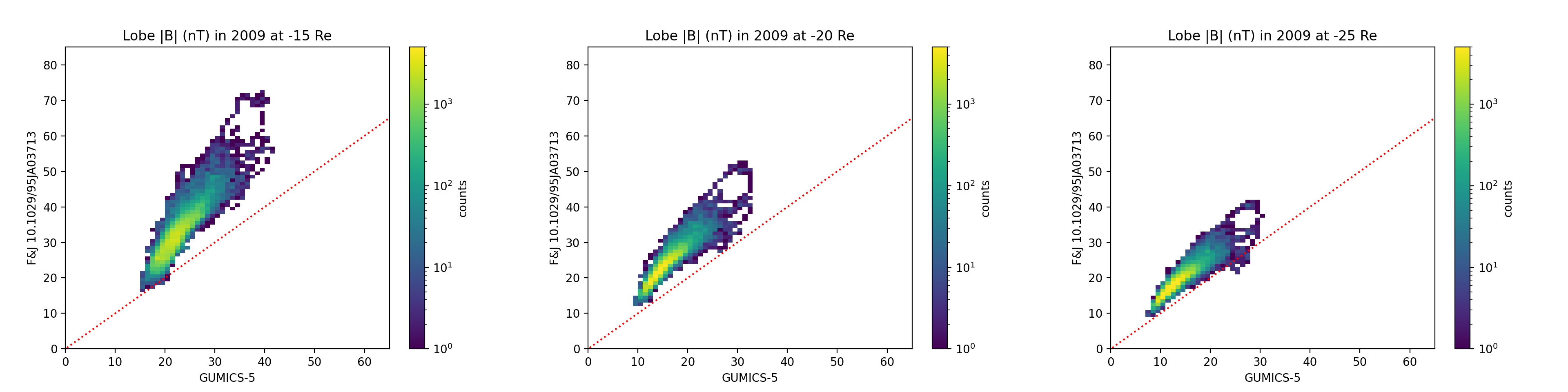}}
\caption{
Same as Figure \ref{fig:lobe2001} but for the minimum between solar cycles 23 and 24.
}
\label{fig:lobe2009}
\end{figure}

Figures \ref{fig:lobe2001} and \ref{fig:lobe2009} compare the magnetic field strength in magnetospheric lobe(s) to the empirical model of \cite{fairfield96}.
From GUMICS-5 we select the maximum magnetic field strength at three distances from Earth along the -X axis at any position in Z direction.

\begin{figure}
\centerline{\includegraphics[width=0.6\paperwidth]{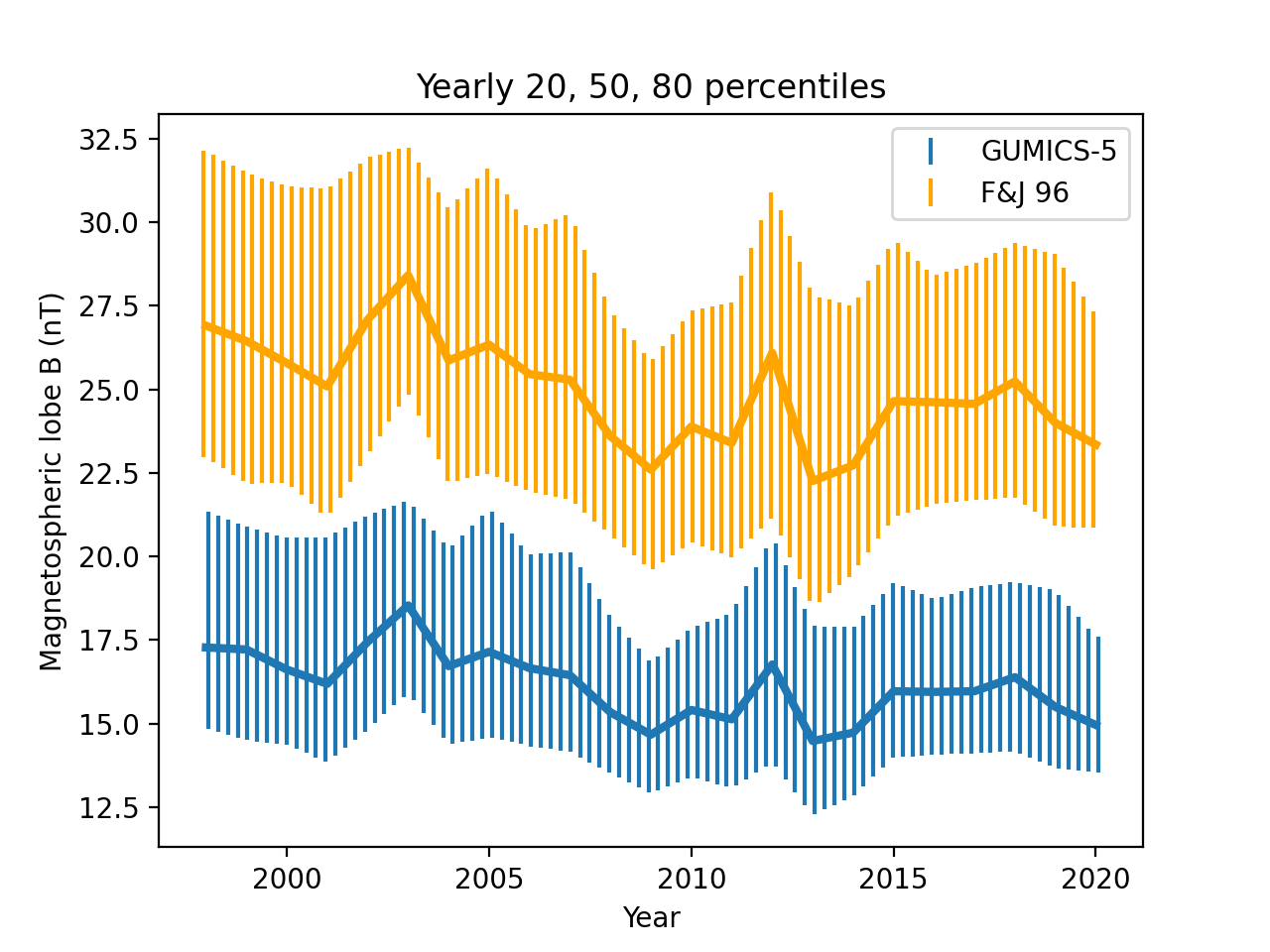}}
\caption{
Statistics of magnetospheric lobe field strength from GUMICS-5 and \cite{fairfield96} at -20 R$_E$ GSE X in Earth's magnetotail.
Solid lines denote yearly medians while tops and bottoms of vertical lines denote 80 and 20 percentiles respectively.
The total number of data points every year is approximately $10^5$.
}
\label{fig:lobeth}
\end{figure}

Figure \ref{fig:lobeth} shows the 20th, 50th and 80th percentiles of GUMICS-5 and empirical magnetospheric lobe field strength for each calendar year between 1998 and 2020 at -20 R$_E$ GSE X in Earth's magnetotail.

GUMICS-5 reproduces the solar cycle time variations of lobe field well when compared to \cite{fairfield96}.
In particular, both approaches show the exceptionally strong lobe fields in 2003 and 2012.
GUMICS-5 underestimates the median field intensity by approximately 10 nT but the underestimation seems to decrease further from Earth.

The empirical model produces values in a much larger range for the solar maximum in 2001 than for the minimum in 2009.
In 2001 the empirical model produces values as low as 0 nT at all studied positions, while GUMICS-5 yields rather fixed minimun values: approximately 15 nT at -15 R$_E$, 8 nT at -20 R$_E$ and 6 nT at -25 R$_E$.
This is not the case in 2009, when both the empirical model and GUMICS-5 yield values above approximately 10 nT.

In \cite{gordeev15}, lobe field strength was underestimated by GUMICS-4, BATS-R-US, LFM, and for large fields by OGGCM as well.

\subsection{Magnetopause standoff distance}

\begin{figure}
\centerline{\includegraphics[width=0.7\paperwidth]{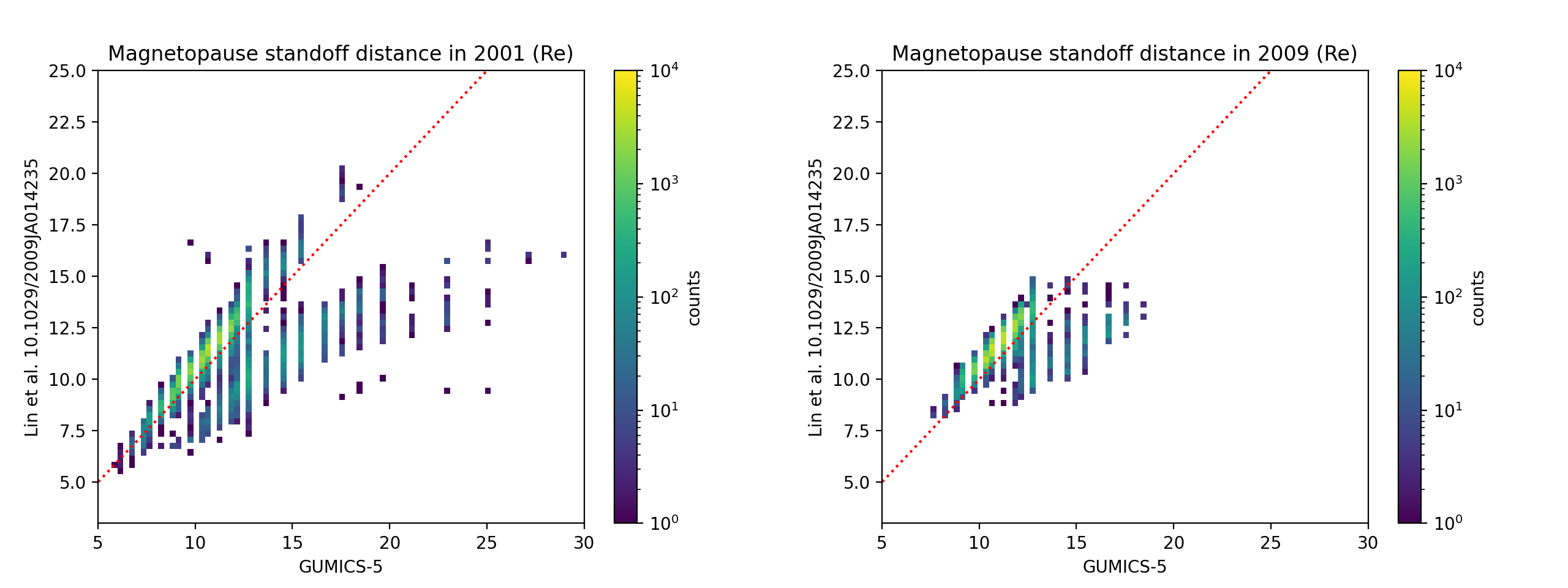}}
\caption{
Magnetopause standoff distance on Sun-Earth line in 2001 during the maximum of solar cycle 23 and in 2009 during the subsequent minimum.
The red line indicates 1:1 correspondence.
}
\label{fig:mpause}
\end{figure}

Figure \ref{fig:mpause} compares the magnetopause standoff distance on the Sun-Earth line to the empirical model of \cite{lin10}.
From GUMICS-5 we select the position on the GSE $X$-axis between 5 and 30 R$_E$ from Earth with maximum $J_y = \nabla \times B_1|_y$, where $B_1$ is the perturbed magnetic field solved by GUMICS-5.
This magnetopause current is a simple and reliable way of detecting the location of the magnetopause along the Sun-Earth line in GUMICS.

\begin{figure}
\centerline{\includegraphics[width=0.6\paperwidth]{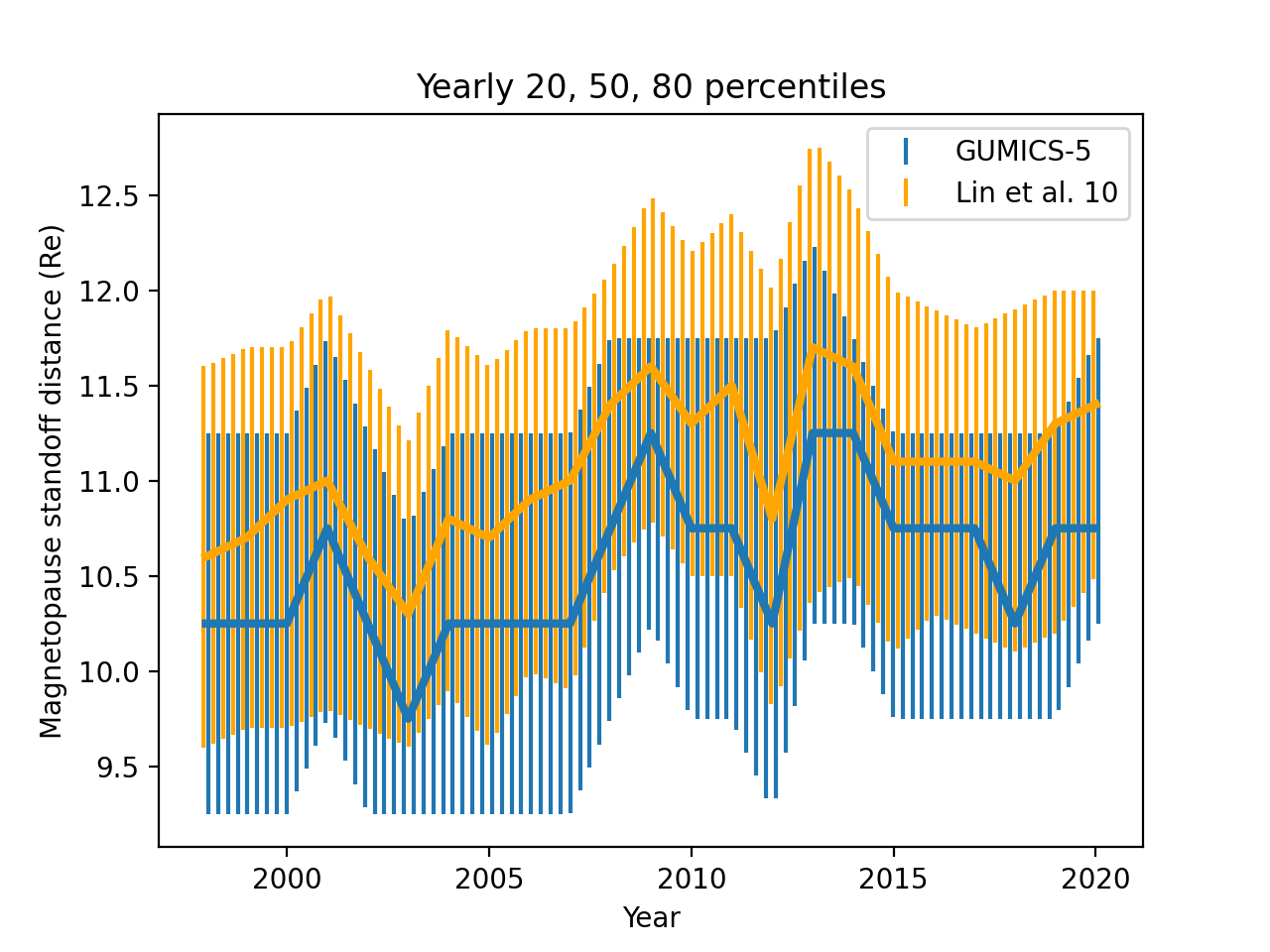}}
\caption{
Yearly percentiles of magnetopause standoff distance from GUMICS-5 and \cite{lin10} in same format as Figure \ref{fig:lobeth}.
}
\label{fig:mpauseth}
\end{figure}

Figure \ref{fig:mpauseth} shows the 20th, 50th and 80th percentiles of GUMICS-5 and empirical magnetopause standoff distance for each calendar year between 1998 and 2020.

GUMICS-5 reproduces the solar cycle time variations in magnetopause standoff distance quite well when compared to \cite{lin10}.
For example the years 2003 and 2012 associated with strong lobe fields (Figure \ref{fig:lobeth}) are local minima in stand-off distance in both GUMICS-5 and the empirical model.
GUMICS-5 underestimates the standoff distance on average by about 0.5 R$_E$, i.e.~by one grid cell, but the two median curves are within their error bars as characterized by the 20th and 80th percentiles.

In 2001 and to some extent in 2009 in Figure \ref{fig:mpause}, the results consist of two distinct populations mostly separated by the sign and magnitude of the solar wind electric field (E$_y$, not shown).
In the larger population IMF $B_z$ is mostly negative and the standoff distance is underestimated by about 0.5 R$_E$.
In the smaller population IMF $B_z$ is positive and, when E$_y$ is larger than about 3 mV/m, GUMICS-5 overestimates the standoff distance by several R$_E$.

The larger population above is close to the results of GUMICS-4 in the lower panel of Figure 4 in \cite{janhunen12}, while in \cite{gordeev15} the standoff distance was also underestimated by OGGCM and LFM and slightly underestimated by GUMICS-4.

\subsection{Plasma sheet pressure}

\begin{figure}
\centerline{\includegraphics[width=0.7\paperwidth]{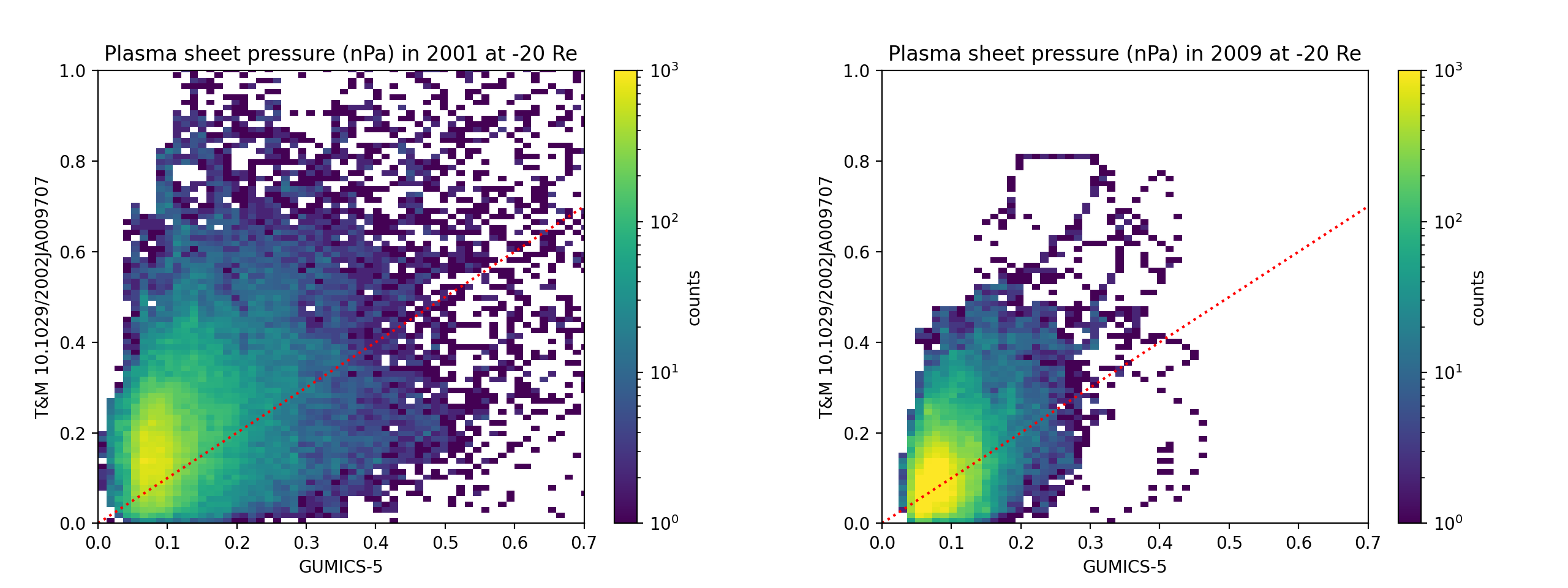}}
\caption{
Plasma sheet pressure at -20 R$_E$ GSE X in the Earth's magnetotail in 2001 during the maximum of solar cycle 23 and in 2009 during the subsequent minimum.
The red line indicates 1:1 correspondence.
}
\label{fig:psheet}
\end{figure}

Figure \ref{fig:psheet} compares plasma sheet pressure to the empirical model of \cite{tsyganenko03}.
From GUMICS-5 we select the maximum thermal pressure within $\pm 10$ R$_E$ in $Y$ and $Z$ dimensions, at three distances from Earth along the GSE $X$ axis.
We only show the results at -20 R$_E$ GSE X as at other distances the results are essentially the same except for a smaller range of values further from Earth and larger range closer to Earth.

\begin{figure}
\centerline{\includegraphics[width=0.6\paperwidth]{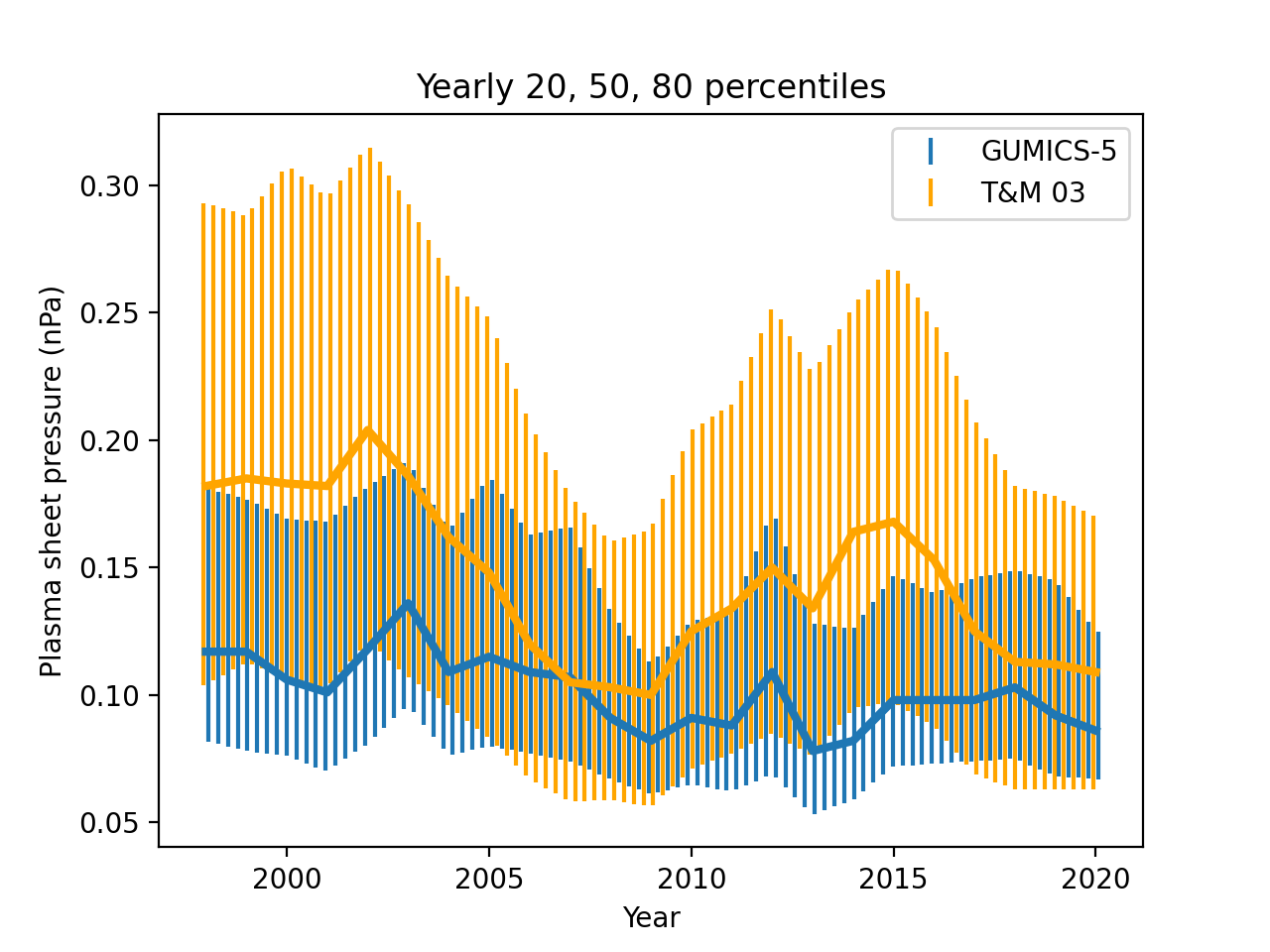}}
\caption{
Yearly percentiles of plasma sheet pressure at -20 R$_E$ GSE X from GUMICS-5 and \cite{tsyganenko03} in same format as Figure \ref{fig:lobeth}.
}
\label{fig:psheeth}
\end{figure}

Figure \ref{fig:psheeth} shows the 20th, 50th and 80th percentiles of GUMICS-5 and empirical plasma sheet pressure at -20 R$_E$ each calendar year between 1998 and 2020.

The modeled and empirically estimated plasma sheet pressures do not appear to be correlated.
This is visible also in the solar cycle time variations, which are less consistent with each other than in the cases of lobe magnetic field and magnetopause standoff distance.
The correlation also does not seem to improve by grouping results based on time of year, $Kp$, hourly average of solar wind $B_z$ nor $V_x$ (not shown).

The behavior of GUMICS-5 is similar to that of GUMICS-4 in \cite{gordeev15}, except that here the spread of input and output values is much wider and correlation is significantly smaller. This is likely due to the stark contrast in coverage of simulation inputs between approximately 200 hours of synthetic versus over 20 years of real solar wind.

\subsection{Cross-polar cap potential}

\begin{figure}
\centerline{\includegraphics[width=0.9\paperwidth]{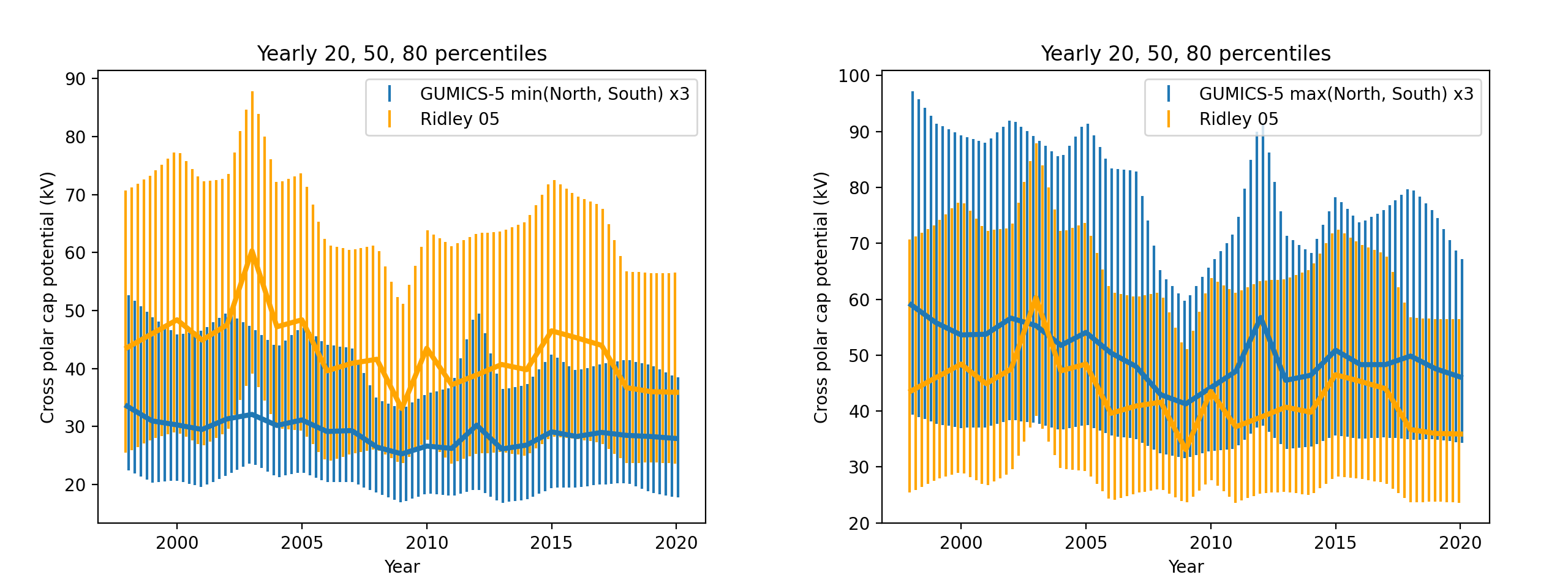}}
\caption{
Yearly percentiles of cross-polar cap potential (CPCP) from GUMICS-5 and \cite{ridley05} in same format as Figure \ref{fig:lobeth}, except that the number of data points is approximately 500 000 per year.
Left shows smaller one of either North or South CPCP from GUMICS-5 while right shows maximum.
Note that GUMICS-5 results have been scaled up by a factor of 3.
}
\label{fig:cpcpth}
\end{figure}

Figure \ref{fig:cpcpth} shows the 20th, 50th and 80th percentiles of cross-polar cap potential from GUMICS-5 and the empirical models of \cite{boyle97,ridley05} on each calendar year between 1998 and 2020.
Using the Alfv\'en Mach number and magnetopause size corrections of \cite{ridley05} seems to minimize the difference between GUMICS-5 and the empirical model, although their effect is small.
Similarly to plasma sheet pressure, there is not much correlation in 5 minute values, nor longer averages, of CPCP between GUMICS-5 and \cite{ridley05} (not shown).

There does not seem to be much correlation between yearly GUMICS-5 and empirical CPCP at any percentile, although both show a decreasing linear trend from 1998 to 2020.
In GUMICS-5 the behaviors of minimums and maximums of both hemispheres' CPCP resemble each other closely.

\subsection{{\it Kp} and GUMICS-5-specific {\it Kp} indices}
\label{sec:kp}

We convert horizontal ionospheric currents from GUMICS-5 to variations of external ground magnetic field ($dB$) using the method of \cite{vanhamaki20}, after which we convert $dB$ to $Kp$ using the same procedure used for real $Kp$:
We convert $dB$ to a local $K$ index using the FMI method described e.g.~in \cite{sucksdorff91, menvielle95} and software available at \href{https://space.fmi.fi/image/software}{space.fmi.fi/image/software}.
We then transform this $K$ index to the standardized index $Ks$ and finally obtain $Kp$ as described in \cite{matzka21}, using the same stations that were used for the real $Kp$.

\begin{figure}
\centerline{\includegraphics[width=0.9\paperwidth]{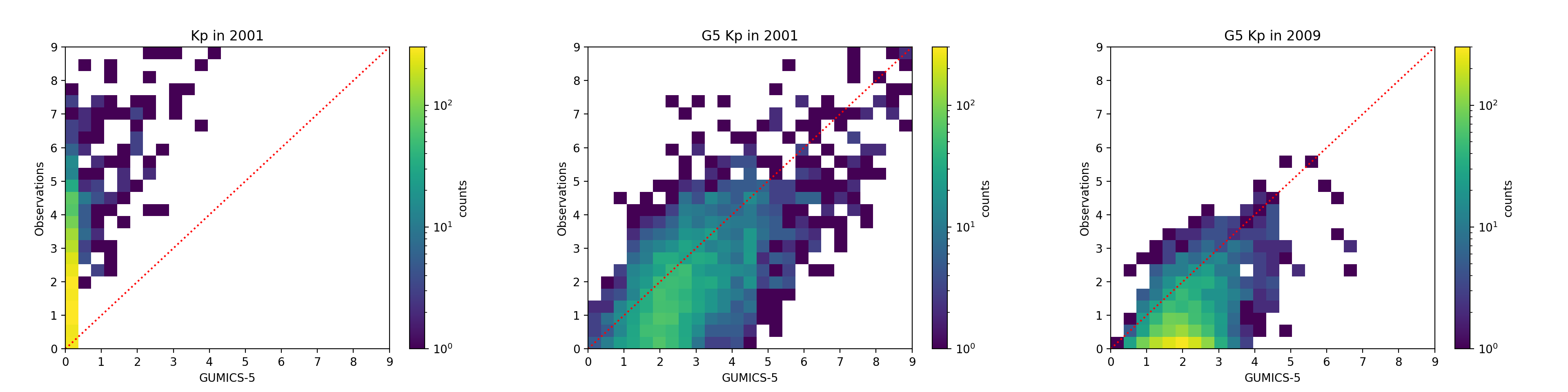}}
\caption{
Left panel:
Real $Kp$ index in 2001 during the maximum of solar cycle 23 obtained from GUMICS-5 and observations.
Middle and right panels:
GUMICS-5-specific $Kp$ index versus real $Kp$ in 2001 and in 2009 during the subsequent solar minimum.
Red line indicates 1:1 correspondence.
}
\label{fig:kp}
\end{figure}

The left panel of Figure \ref{fig:kp} compares the $Kp$ index obtained from GUMICS-5 to the measured one.
Due to the significantly smaller amplitude of GUMICS-5-derived ground $dB$ compared to measurements, GUMICS-5 severely underestimates $Kp$.
Hence, we derive GUMICS-5-specific $K$ indices for all stations used in $Kp$ by adjusting the station-specific K9 threshold such that the frequency of $K \geq 5$ is approximately 5 \% (\cite{matzka21} and references therein).
After deriving the GUMICS-5-specific $K$ index, the same procedure as above is used to compute the GUMICS-5-specific $Kp$ index.
This brings the dynamic range of the GUMICS-5-specific $Kp$ to the same level as the observed one and is compared to real $Kp$ in middle and right panels of Figure \ref{fig:kp}.

In Figure \ref{fig:kp} the minimum value of GUMICS-5-specific $Kp$ seems to be mostly between 1 and 2, contrary to measurements, especially during solar minimum in 2009 when real $Kp < 1$ is seen most often.
During solar maximum in 2001, real $Kp$ values are divided somewhat evenly in the range from 0 to 2, while a value of 2 is most often seen in GUMICS-5-specific $Kp$.

\begin{figure}
\centerline{\includegraphics[width=0.6\paperwidth]{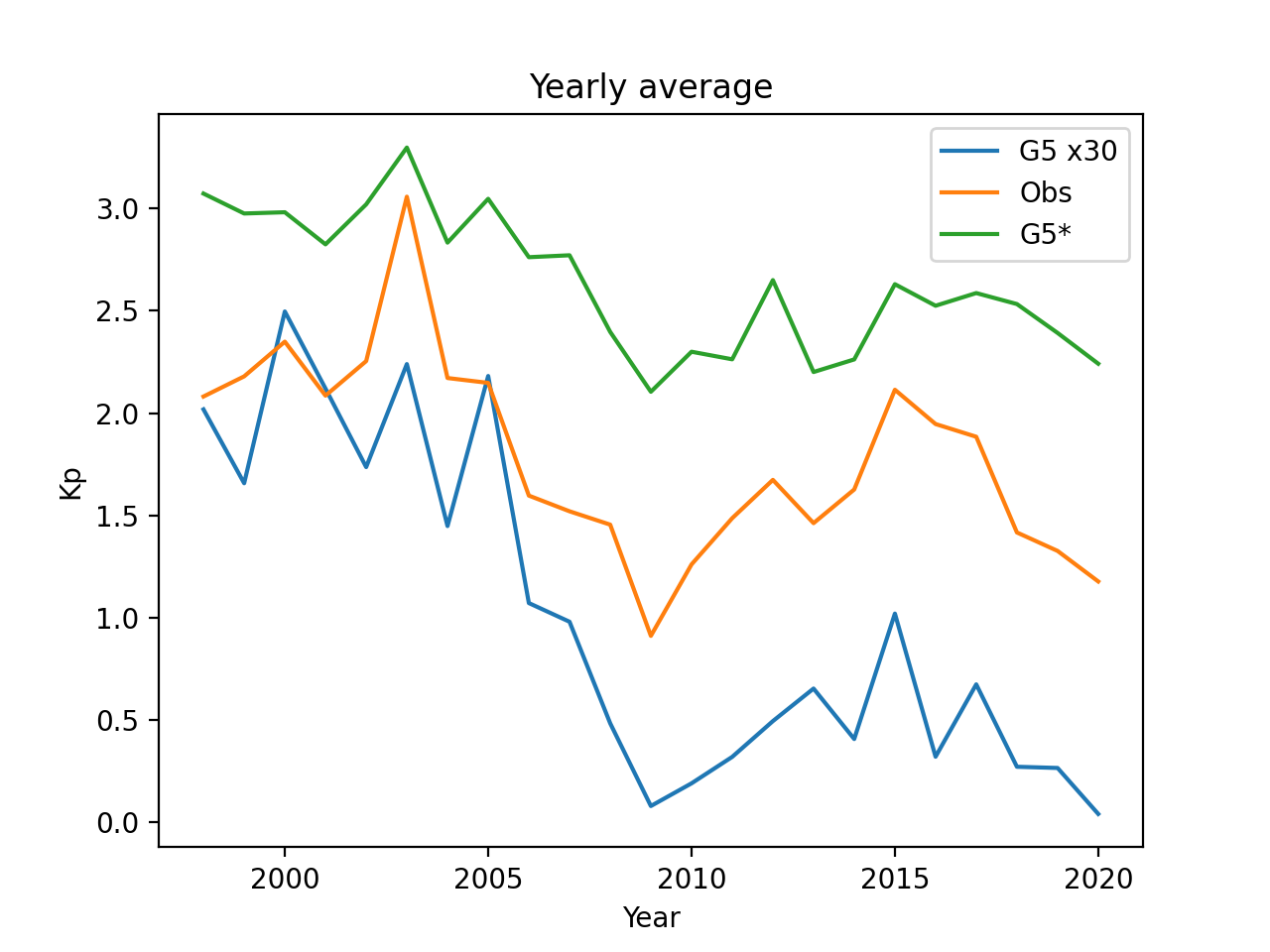}}
\caption{
Yearly averages of real $Kp$ (Obs), GUMICS-5-specific $Kp$ (G5\textasteriskcentered) and real $Kp$ derived from GUMICS-5 scaled up by a factor of 30 (G5 x30).
}
\label{fig:kpavg}
\end{figure}

\begin{figure}
\centerline{\includegraphics[width=0.6\paperwidth]{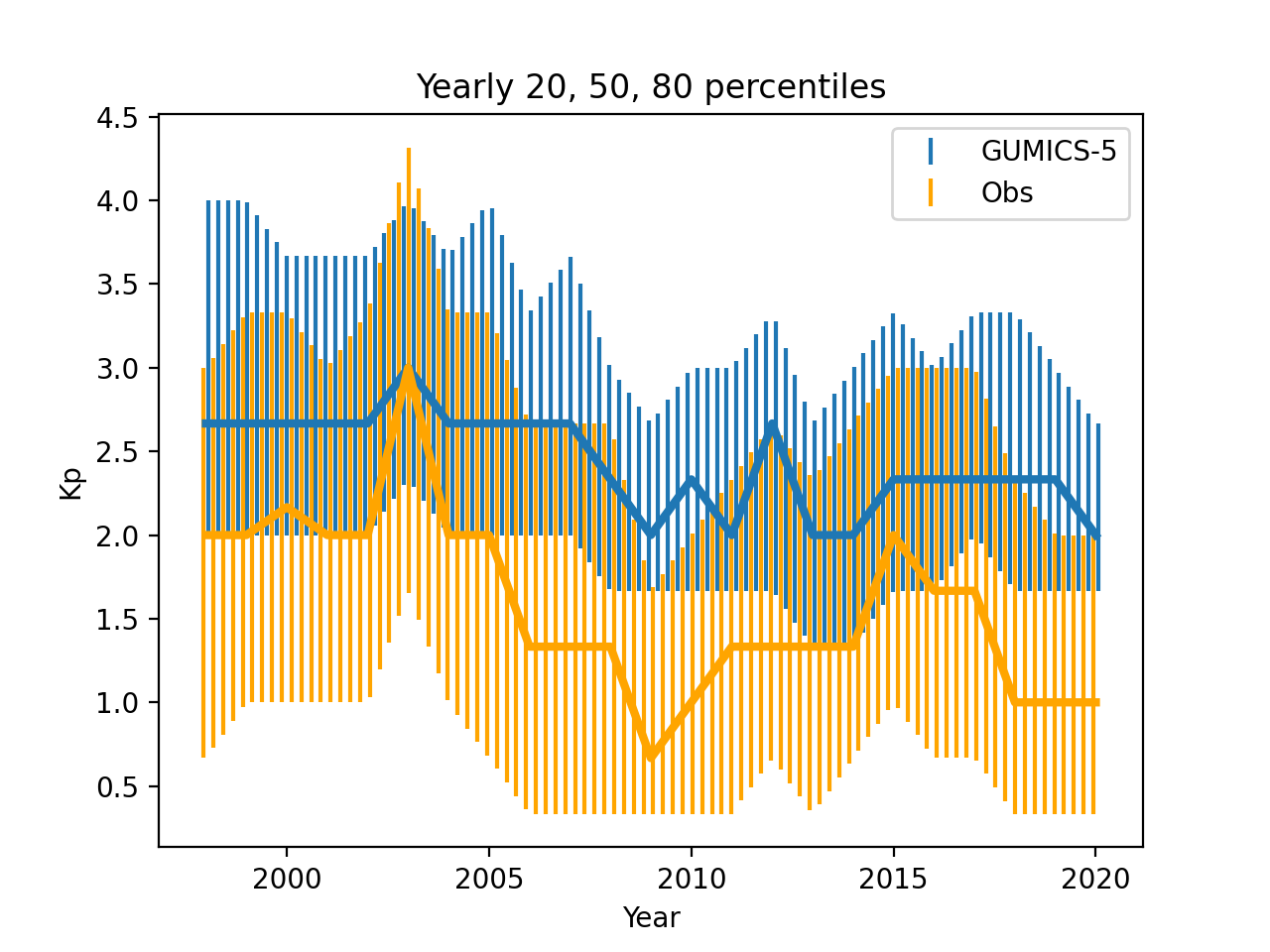}}
\caption{
Yearly percentiles of real and GUMICS-5-specific $Kp$ indices in same format as Figure \ref{fig:lobeth}, except that the number of data points is approximately 2900 per year.
}
\label{fig:g5kpth}
\end{figure}

Figure \ref{fig:kpavg} shows the yearly averages of real $Kp$, real $Kp$ calculated from GUMICS-5 and GUMICS-5-specific $Kp$ while Figure \ref{fig:g5kpth} shows the 20, 50 and 80th percentiles of real and GUMICS-5-specific $Kp$ each calendar year between 1998 and 2020.

In \cite{honkonen18}, horizontal ionospheric currents accounted for approximately 90\% of modeled ground $dB$.
However the effects of ring current and radiation belts on ground $dB$ were not studied, hence their contribution to $dB$ produced by GMHD models is less well known.
As the ring current and radiation belts are neglected also here, our $Kp$ estimates are more dependent on high-latitude activity than the observed $Kp$ is.

\subsection{Auroral electrojet index}
\label{sec:ae}

\begin{figure}
\centerline{\includegraphics[width=0.7\paperwidth]{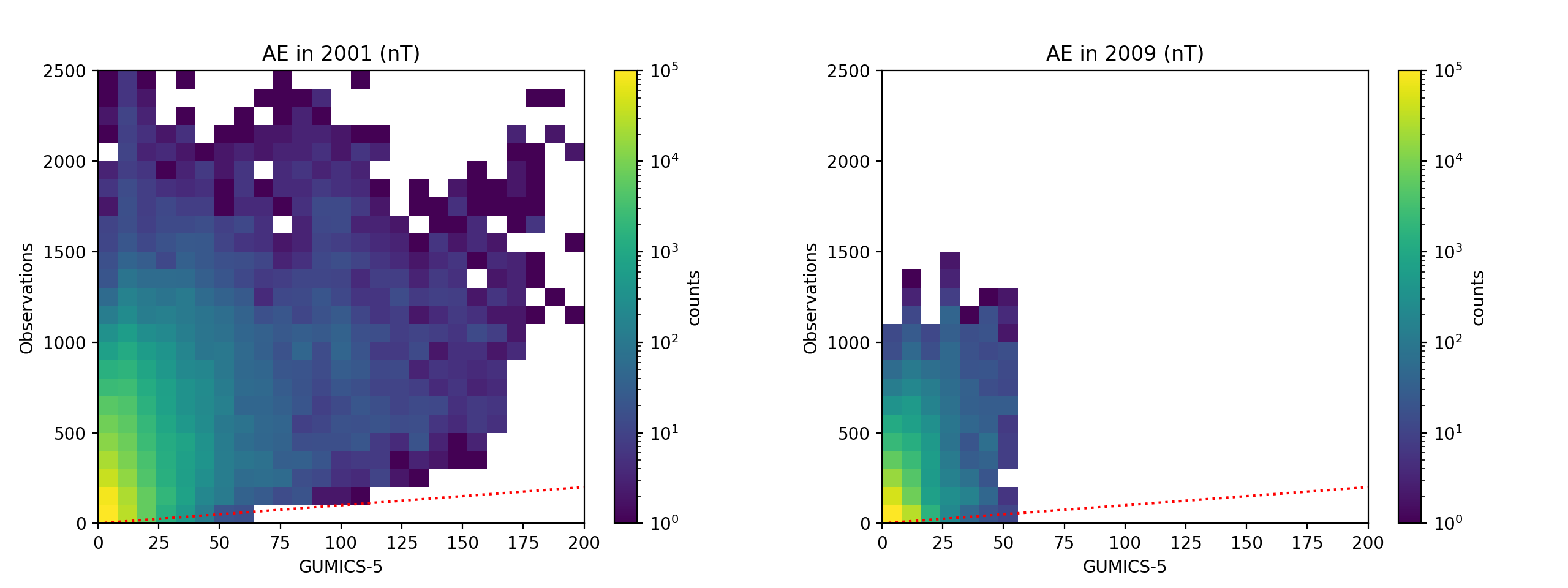}}
\caption{
Auroral electrojet index in 2001 during the maximum of solar cycle 23 and in 2009 during the subsequent minimum.
The red line indicates 1:1 correspondence.
}
\label{fig:ae}
\end{figure}

\begin{figure}
\centerline{\includegraphics[width=0.6\paperwidth]{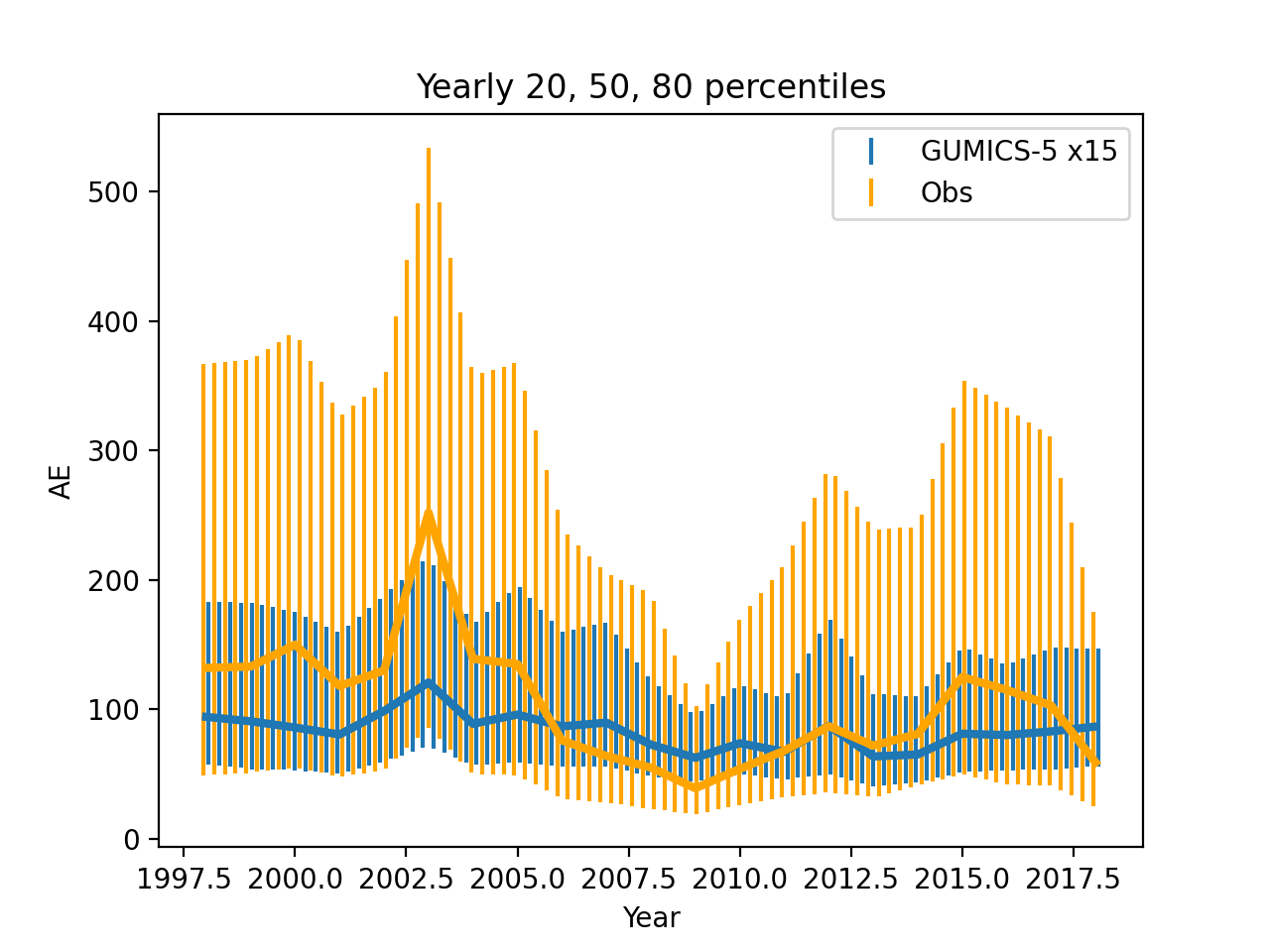}}
\caption{
Yearly percentiles of real and GUMICS-5-derived auroral electrojet ($A_E$) index in same format as Figure \ref{fig:lobeth}.
Note that GUMICS-5 results have been scaled up by a factor of 15.
The measured $A_E$ index is available only up to 2018-02-28.
}
\label{fig:aeth}
\end{figure}

Figure \ref{fig:ae} compares the auroral electrojet index ($A_E$) obtained from GUMICS-5 to the real one and Figure \ref{fig:aeth} shows their 20th, 50th and 80th percentiles
each calendar year between 1998 and 2018 when observed $A_E$ is available.
The method for deriving $A_E$ is described by \cite{davis66}.
Determining $A_E$ involves removal of the quiet-day baseline, which is defined as the average of the horizontal magnetic field component over the 5 international quietest days of every month.
Quiet-day determination is based on the $Kp$ index and we find relatively small differences in $A_L$ and $A_U$ indices produced by GUMICS-5 (not shown) whether we use observed quiet days or quiet days determined from GUMICS-5-specific $Kp$.
The $A_E = A_U - A_L$ index is not affected by the determined baseline.

There are some similarities in the yearly percentiles of $A_E$ between measurements and GUMICS-5 but their correlation is not very good.
The worse correlation of GUMICS-5 versus $A_E$ compared to $Kp$ is likely due to the fact that $A_E$ reflects more the effects of e.g.~substorms, whose formation is less directly driven by solar wind and hence is more difficult to capture using GMHD models.
The higher time resolution of $A_E$ likely also contributes to worse correlation with GUMICS-5 compared to $Kp$.

\section{Discussion}
\label{sec:discussion}

Previous comparisons of GUMICS results with empirical models have shown that the simulation underestimates the intensity of magnetospheric and ionospheric activity (\cite{juusola14}, \cite{gordeev13}).
However, like shown in \cite{palmroth05}, the time evolution of substorm activity as described by the simulation is in reasonable agreement with observations.
In this study our main objective is to determine whether GUMICS-5 can reproduce the time evolution of the magnetosphere and ionosphere over one or more solar cycles. 
When considering the yearly statistic this is indeed the case for the magnetospheric lobe magnetic field strength, the magnetopause standoff distance and the $Kp$ index, although GUMICS-5 requires a constant offset or scaling factor for best fit to observations and empirical models.
The auroral electrojet index and plasma sheet pressure are not reproduced as well and GUMICS-5 does not seem to be able to reproduce the solar cycle effects of cross-polar cap potential observed in empirical models.

\subsection{Comparison to GUMICS-4 and other GMHD models}

The strength of magnetospheric lobe field is modeled similarly by both GUMICS-5, GUMICS-4 and other GMHD models.
In \cite{gordeev15} lobe field was underestimated by GUMICS-4, BATS-R-US and LFM, and for large fields by OGGCM as well.

The behavior of GUMICS-5 w.r.t~plasma sheet pressure is also similar to GUMICS-4 in \cite{gordeev15}, except that here the spread of input solar wind and output pressure values is much wider and correlation is significantly smaller.
This is likely due to the fact that we have simulated an approximately three orders or magnitude longer time interval than was simulated in \cite{gordeev15}.

The magnetopause standoff distance is modeled similarly by GUMICS-5 and GUMICS-4, when comparing to the lower panel of Figure 4 in \cite{janhunen12}.
More specifically the behavior of both models seems similar at least when IMF $B_z$ is mostly southward, although this could due to the much shorter amount of simulations performed with GUMICS-4.
In \cite{gordeev15} magnetopause standoff distance was also underestimated by OGGCM and LFM and slightly underestimated by GUMICS-4.

\subsection{Sources of modeling error}
\label{sec:errors}

The poor performance of GUMICS-5 in some of the comparisons can at least partly be assigned to the following causes. 

Due to limitations in computational and storage capacity we changed the highest magnetospheric resolution from 0.25 R$_E$, used mostly with GUMICS-4, to 0.5 R$_E$ which had been used in early GUMICS-4 simulations.
In future studies, with larger computational and storage resources available and with further optimization of GUMICS-5, we hope to use the more standard value of 0.25 R$_E$.
Lower magnetospheric resolution increases numerical diffusion which might cause GUMICS-5 to e.g.~underestimate the maximum magnetospheric lobe field strength.
Lower resolution also decreases the magnitude of field-aligned currents, causing e.g.~lower horizontal ionospheric currents.

GUMICS-5 lacks an inner magnetosphere model which is essential for reproducing many low-latitude phenomena \cite{liemohn18}.
Incorporating an inner magnetosphere into GUMICS would require a large effort and likely should be done in the context of overall modernization of GUMICS' solvers.

The poor agreement of plasma sheet pressure with the empirical model might be explained by the lack of an ionospheric outflow model in GUMICS-5.
Adding solar activity-dependent outflow might improve plasma sheet pressure results considerably over solar cycle time scales, while adding ionospheric activity-dependent outflow could improve the results over hourly and/or daily time scales.

We have set $B_x=0$ in the solar wind, in order to have a divergence-free magnetic field at the solar wind boundary.
In the future, a straightforward solution could be to introduce a slow, e.g.~1 mHz at most, global variation in the background $B_x$ within the simulated volume, in order to better take into account a changing IMF $B_x$ while maintaining a divergence-free field.
This could perhaps be calculated automatically from the solar wind magnetic field already given as input to GUMICS-5.

Similarly to most simulations carried out with GUMICS-4 \cite{janhunen12}, we use a constant solar EUV flux that is parametrized by solar radio F10.7 cm flux.
We expect the effect of changing F10.7 flux to be small based on \cite{gordeev13}.

We have not omitted simulations with gaps in solar wind data even though sufficiently long gaps could cause GUMICS-5 to miss certain short-term effects in magnetosphere or ionosphere.
Most likely this does not impact results presented here but it should be kept in mind when analysing e.g.~individual storms.

The relatively low magnetospheric resolution in GUMICS-5 used here, as well as the first-order MHD solvers, likely also contribute to the low correlation between GUMICS-5 and $A_E$ compared to the correlation with $Kp$.
We also neglect the effect of ground conductivity when deriving geomagnetic indices.
Its effect on the largest horizontal ground magnetic fields can be a few tens of percent \cite{juusola20} but such a relatively small error will likely not change our results significantly.

\subsection{Future optimizations}

One or more of the following optimizations will likely be required in order to e.g.~increase the highest magnetospheric resolution from 0.5 to 0.25 R$_E$ of the simulations presented here.

We have already reduced the size of magnetospheric output files by approximately 50 \% in post-processing by omitting the face-average background magnetic field stored for all faces of magnetospheric cells.
This is used by the MHD solvers but is not necessary for e.g.~obtaining the results presented here.
The ionospheric files are about 1/100 of the size of the original magnetospheric files and most likely will not require optimization.

A further 50\% reduction in size can be obtained by converting results from 64 bit to 32 bit floating point representation, either in GUMICS-5 or as an additional post-processing step.
Significant further savings would require e.g. leaving out some parts of the magnetosphere most of the time, which would affect future analysis and the ability to restart the simulation.
In that case, existing analysis methods could be applied directly after the simulation, before discarding parts of magnetospheric results, thereby saving space in the long run without affecting comparisons to previous results.

Optimizing CPU performance of GUMICS-5 is currently not as important as lowering the storage requirements of results, but there is at least one significant potential optimization.
Currently in GUMICS-5 the magnetosphere is parallelized only using MPI, but taking advantage of OpenMP threading could further increase processing speed and reduce memory requirements.

\section{Conclusions}
\label{sec:conclusions}

We present our approach to modeling over 20 years of the solar wind-magnetosphere-ionosphere system using version 5 of the Grand Unified Magnetosphere-Ionosphere Coupling Simulation (GUMICS-5).

We compare the simulation results to several empirical models of the magnetosphere and ionosphere, and to geomagnetic indices derived from ground magnetic field measurements.
GUMICS-5 reproduces observed solar cycle trends in magnetopause stand-off distance and magnetospheric lobe field strength, while consistency in plasma sheet pressure and ionospheric cross-polar cap potential is lower.
Comparisons with geomagnetic indices show better results for $Kp$ index than for $A_E$ index.

We run over 8000 1-day simulations, save the magnetospheric results every 5 minutes and ionospheric results every minute, producing over 2 and 10 million files respectively that required over 100 TB of disk space in total.
We describe the simulation setup and challenges related to such scale of global magnetohydrodynamic modeling.

Our extensive results can serve e.g.~as a foundation for a combined physics-based and black-box approach to real-time prediction of near-Earth space, or as input to other physics-based models of the inner magnetosphere, upper and middle atmosphere, etc.

\section{Acknowledgements}
We thank Ari Viljanen for insightful discussions and the Finnish Center for Scientific Computing for providing help with and managing the Puhti supercomputer, Allas object storage system and Fairdata service.
We are grateful to NASA National Space Science Data Center, the Space Physics Data Facility, and the ACE Principal Investigator, Edward C. Stone of the California Institute of Technology for providing access to ACE data.
We are also grateful to GFZ German Research Centre for Geosciences for $Kp$ and quiet days data, and to Kyoto World Data Center for Geomagnetism for $A_E$ data \cite{ae}.

\bibliographystyle{abbrv}
\bibliography{references.bib}

\end{document}